\documentclass[9pt,conference]{IEEEtran}
\usepackage{amssymb,amsthm,amsmath,array}
\usepackage{graphicx}
\usepackage[caption=false,font=footnotesize]{subfig}
\usepackage{xspace}
\usepackage[sort&compress, numbers]{natbib}
\usepackage{stmaryrd}
\usepackage{xcolor}
\usepackage{mathtools}
\usepackage{float}
\usepackage{textcomp}
\usepackage{amsmath,amssymb,amsfonts}
\usepackage{algorithm}
\usepackage{algorithmic}
 \usepackage{multirow}
\usepackage{bigstrut}
 \graphicspath{{./fig/}}
 
\usepackage{booktabs} 
\DeclareMathOperator*{\argmax}{ \operatorname{argmax}}

\usepackage{textcomp}
\usepackage{xcolor}
\def\BibTeX{{\rm B\kern-.05em{\sc i\kern-.025em b}\kern-.08em
    T\kern-.1667em\lower.7ex\hbox{E}\kern-.125emX}}

\def\bsw{{\boldsymbol{w}}}
\def\bsx{{\boldsymbol{x}}}

\newcounter{algo}
\renewcommand{\thealgo}{\arabic{algo}}

\begin{document}
\title{Bayesian Based Unrolling for Reconstruction and Super-resolution of Single-Photon Lidar Systems }

\author{\IEEEauthorblockN{
        Abderrahim Halimi\IEEEauthorrefmark{1} \thanks{This work was supported by the UK Royal Academy of Engineering under the Research Fellowship Scheme (RF/201718/17128) and EPSRC Grants EP/T00097X/1,EP/S026428/1.}, 
        Jakeoung Koo\IEEEauthorrefmark{4}, and 
        Stephen McLaughlin\IEEEauthorrefmark{1}
    }
    \IEEEauthorblockA{
        \IEEEauthorrefmark{1} School of Engineering and Physical Sciences, Heriot-Watt University, Edinburgh UK.\\
        \IEEEauthorrefmark{4}  Gachon University, South Korea.  }
} \vspace{-0.3cm}
\maketitle

\begin{abstract}
Deploying 3D single-photon Lidar imaging in real
world applications faces several challenges due to  imaging in high noise environments and with sensors having limited resolution. This paper presents a deep learning algorithm based on unrolling a Bayesian model for the reconstruction and super-resolution of 3D single-photon Lidar. The resulting algorithm benefits from the advantages of both statistical and learning based frameworks, providing best estimates with improved network interpretability. 
Compared to existing learning-based solutions, the proposed architecture requires a reduced number of trainable parameters, is more robust to noise and mismodelling of the system impulse response function, and provides richer information about the
estimates including uncertainty measures. Results on synthetic
and real data show competitive results regarding the quality of
the inference and computational complexity when compared to
state-of-the-art algorithms.  

This short paper is based on contributions published in \cite{koo2022bayesian} and \cite{halimi2021robust}.
\end{abstract} \vspace{-0.3cm}

\section{Introduction} \vspace{-0.1cm}
3D Single-photon  Lidar imaging operates by sending light pulses and collecting the reflected photons from a target. Rapid or long range imaging result in the detection of a reduced number of photons. In addition, imaging in bright conditions or through obscurants can increase the background noise affecting the measurement. Several methods have been proposed to deal with these challenges by exploiting multiscale information, spatial correlation and data statistics and we can group them into  statistical based methods \cite{rapp2017few,Tachella_NC_2019,Tachella_Siam2019,halimi2021robust} and deep learning methods \cite{lindell2018singlephoton,peng2020photonefficient,ruget2021robust}.   In this paper, we unroll a statistical model into a deep learning architecture, hence providing depth and uncertainty estimates with improved network interpretability. We validate the approach on depth maps reconstruction and super-resolution under extreme imaging conditions.
 \vspace{-0.2cm}

\section{Underlying Bayesian algorithm}  \vspace{-0.1cm}
A Lidar system obtains a histogram of counts $y_{n,t}$ at the $n$-th pixel and the $t$-th time bin which follows a Poisson likelihood $y_{n, t} \sim \mathcal{P}\left[r_{n} \, g \left(t-d_{n}\right)+b_{n,t}\right]$, where $r_{n}, d_n$ denotes the target reflectivity and depth, $g(.)$ the system impulse response and $b_{n,t}$ the background noise.  The multiscale approximate Bayesian model has been proposed in \cite{halimi2021robust} to deal with noisy data. It introduces few approximations to obtain a simplified likelihood  given by
\begin{equation} \label{eq:pynl2}
\begin{aligned}
p\left( \boldsymbol y_{n}^{(\ell)} \mid r_{n}^{(\ell)}, d_{n}^{(\ell)}\right) & \propto \mathcal{G}\left(r_{n}^{(\ell)} ; 1+\bar{s}_{n}^{(\ell)}, 1\right) Q \left( \boldsymbol{y}_{n}^{(\ell)}\right) \\
&{\times}  \mathcal{N}\left(d_{n}^{(\ell)} ; d_{n}^{\mathrm{ML}(\ell)}, \bar{\sigma}^{2(\ell)}\right),
\end{aligned}
\end{equation}
where  $\boldsymbol y_n^{(\ell)}$ with $\ell \in \{1, 2, ..., L\}$ denotes the $\ell$th downsampled histogram of counts, $Q(.), \bar{s}_{n}^{(\ell)}, \bar{\sigma}^{2(\ell)}$ relate  to the number of detected photons and ML stands for maximum likelihood. 
A Bayesian model is introduced  to estimate a single depth map $\bsx$ and its uncertainty $\boldsymbol{\epsilon}$ by assigning  them Laplace $\mathcal L$ and inverse gamma $\mathcal{I} \mathcal{G}$ prior distributions respectively, as follows:   

  \begin{small}
\begin{equation} \label{eq:xndn}
\begin{array}{c}
x_{n} \mid d_{\nu_{n}}^{(1, \cdots, L)}, w_{\nu_{n}, n}^{(1, \cdots, L)}, \epsilon_{n} \sim  
\prod_{n^{\prime} \in \nu_{n}}\left[\prod_{\ell=1}^{L} \mathcal{L}\left(x_{n} ; d_{n^{\prime}}^{(\ell)}, \epsilon_{n} / w_{n^{\prime}, n}^{(\ell)}\right)\right] \\
\boldsymbol \epsilon \sim \prod_n \mathcal{I} \mathcal{G}\left(\epsilon_{n} ; \alpha_d, \beta_d \right)
\end{array},
\end{equation}
\end{small}
where $\boldsymbol \epsilon=\left( \epsilon_1, \cdots,  \epsilon_N \right)$, with $\epsilon_n$ the variance of the depth $x_n$,  $\nu_n$ is a spatial neighborhood around the pixel $n$; ;   $w_{n^{\prime}, n}^{(\ell)}$ the pre-defined weights to guide the correlation between multiscale depths $d_{n}^{(\ell)}$ and the latent variable $x_n$ and  $\alpha_d$ and $\beta_d$ are user set positive hyperparameters.  The estimation is then performed by maximizing the resulting posterior distribution in \eqref{eq:Posterior} as described in Algo. \ref{alg1}. 
\begin{equation} \label{eq:Posterior}
p\,(\boldsymbol{x}, \boldsymbol{\epsilon}, \boldsymbol{D} \mid \boldsymbol{Y}, \boldsymbol{W}) \propto \, p\,(\boldsymbol{Y} \mid \boldsymbol{D}) \, p\,(\boldsymbol{x}, \boldsymbol{D} \mid \boldsymbol{\epsilon}, \boldsymbol{W}) \, p \,(\boldsymbol{\epsilon}).
\end{equation}

\begin{algorithm}[ht]
\caption{Iterative Bayesian algorithm~\cite{halimi2021robust}} \label{alg1}
\begin{algorithmic}[1]
       \STATE \underline{Input}: Lidar data $\boldsymbol Y$, the number of scales $L$
       \STATE Construct downsampled histograms $\boldsymbol Y^{(1,...,L)}$
       \STATE Compute initial multiscale depths $\boldsymbol{d}^{\mathrm{ML}(1,...,L)}$
       \STATE Compute the guidance weights $\boldsymbol W$
       \WHILE{not converge}
		\STATE Update the variable $\boldsymbol x$ by maximizing \eqref{eq:Posterior}
		\STATE Update the multiscale depths $\boldsymbol d^{(1,\cdots, L)}$ by   maximizing \eqref{eq:Posterior} 
		\STATE Update the uncertainty information by  maximizing \eqref{eq:Posterior} 
               \STATE \textbf{break} if the convergence criteria are satisfied
       \ENDWHILE
   
	\STATE \underline{Output}: $\boldsymbol x, \boldsymbol \epsilon$  
\end{algorithmic}
\end{algorithm} 

\def\fw{420pt}
\begin{figure*}[t]
\centering
\includegraphics[width=\fw]{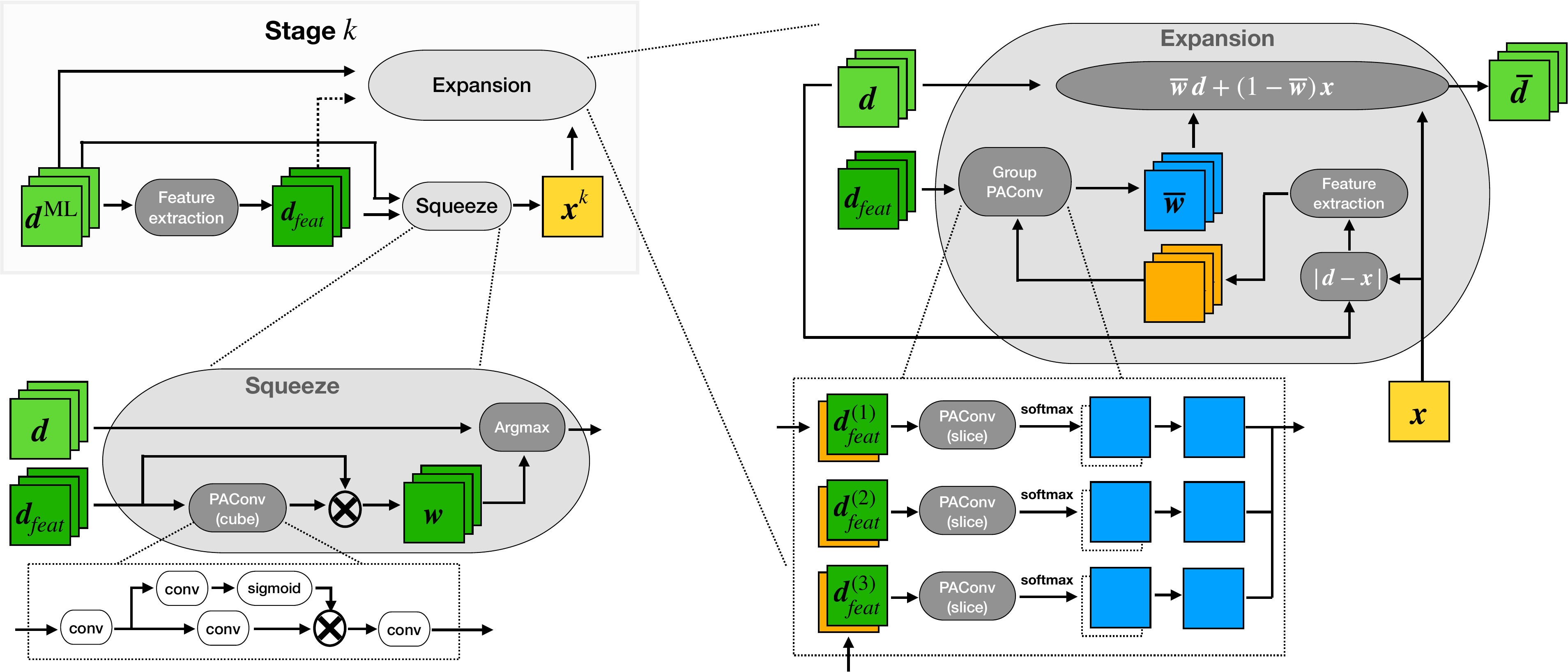}
\caption{The proposed network architecture for one stage $k$ when $L=3$. Each stage consists of three main blocks: feature extraction, squeeze block and expansion block. All the feature extraction layers consist of three convolution layers.}
\label{fig:network}
\end{figure*}

\def\fh{220pt}
\begin{figure}[ht]
\centering
\includegraphics[width=8.5cm]{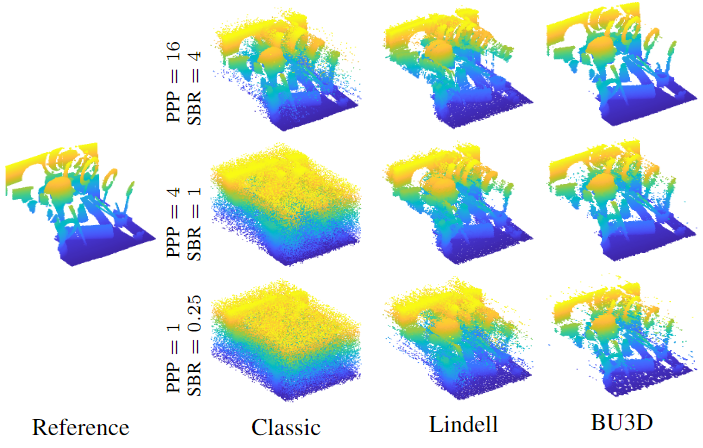}
\caption{Point cloud representation of reconstruction results on the Art scene. The first column shows the reference point cloud. }
\label{fig:Results_Denoising_Art_Scene}
\end{figure}

\def\fh{220pt}
\begin{figure}[ht]
\centering
\includegraphics[width=6.5cm]{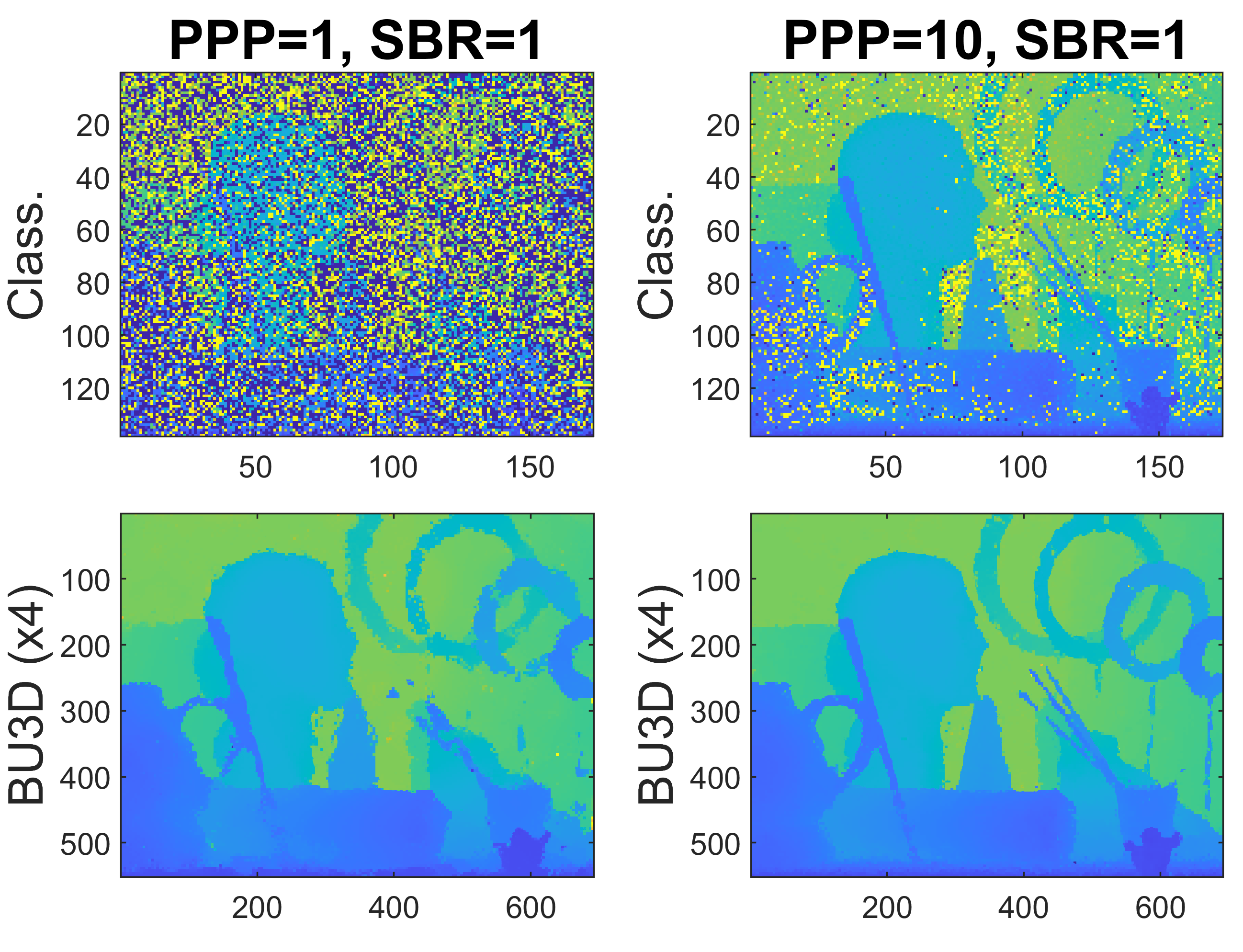}
\caption{Estimated depth maps by BU3D with an upscaling factor $r=4$ on the Art scene. }
\label{fig:Results_SR}
\end{figure}

\section{Unrolling method} 
Our network unrolls the steps of Algo. \ref{alg1} into $K$ stages having the same structure except for the last one, and the weights are not shared among stages. As shown in Fig.~\ref{fig:network}, each stage inputs a set of multiscale depths $\boldsymbol d$ and consists of feature extraction, the squeeze block and the expansion block. The features of $\boldsymbol d$ are extracted by three convolution layers with $3 \times 3$ filters. Throughout the network, all the convolutional layers use the $3 \times 3$ filters with LeakyReLU activation except for PAConv shown in Fig.~\ref{fig:network}.

\textbf{Squeeze block.} The obtained features are fed into the squeeze block which mimics the weighted median filtering in~\cite{halimi2021robust}. The squeeze block relies on an attention layer named PAConv to compute attention weights $\{ w_n^{(\ell)} \}$ that indicate the importance of each scale within a given pixel. The squeezed depth $\boldsymbol x$ is obtained by selecting one scale for each pixel to approximate the weighted median operator as follows
%
\begin{equation} \label{eq:argmax}
x_n = d_n^{(\ell ')}, \quad \ell' = \argmax_{\ell \in \{1,\cdots,L \}} w_n^{(\ell)}.
\end{equation}

\textbf{Expansion block.} The squeezed depth $\boldsymbol x$, the multiscale depths $\boldsymbol d$ and its features are fed into the expansion block. This block corresponds to the generalized shrinkage operator in~\cite{halimi2021robust} and its goal is to refine the multiscale depths. To emphasize the relative difference, the block computes $|\boldsymbol d^{(\ell)} - \boldsymbol x|, \forall \ell$ whose features are used to compute another attention weights. Unlike the squeeze block, the expansion block computes attention weights slice by slice, normalizing between 0 and 1. These normalized weights $\boldsymbol {\overline w}$ are used to compute the refined multiscale depths as $\boldsymbol{\overline d}^{(\ell)} = \boldsymbol{\overline w}^{(\ell)} \boldsymbol d^{(\ell)} + (1 - \boldsymbol{\overline w}^{(\ell)}) \boldsymbol x$.

The refined multiscale depths $\boldsymbol{\overline d}^{(\ell)}, \forall \ell$ are again used as an input of the next stage. The last stage considers the squeezed depth as  the final estimated depth and has no expansion block.

\textbf{Extension for super-resolution} 
The network can be extended to perform depth map super-resolution by a factor $r$. This is achieved by introducing few changes in the squeeze and expansion blocs. In the squeeze bloc, we propose to upsample the weights $\bsw$ using the ESPCN algorithm in  \cite{shi2016real} resulting in an upsampled $\bsx$. In the expansion bloc, the differences are now computed by $| d^{(\ell)}_n -  x_{\psi_n}|, \forall \ell$ 
where the $n$th pixel of $d^{(\ell)}_n$ is compared to the  $r \times r$ patch of $x$ denoted by ${\psi_n}$. The   PAConv (slice) are  updated accordingly to account for the change in dimensions.

\section{Results on simulated data}
We evaluate the algorithm on simulated data. We  train the model using 9 scenes from the Middlebury stereo dataset~\cite{hirschmuller2007evaluation} (with image sizes $555{\times}650$) and   21 scenes from the Sintel stereo dataset~\cite{butler2012naturalistic} (with   image sizes $436{\times}1024$). The test is performed on the Art scene. The histograms of counts are generated based on the Poisson observation model, with different levels of  average number of  Photons-Per-Pixel ($\text{PPP}=\frac{1}{N}\sum_{n=1}^N \left(r_n + b_{n} T\right)$); and  average Signal-to-Background Ratio $\text{SBR}=\frac{\sum_{n=1}^N r_n} { \sum_{n=1}^N  b_{n} T }$. Fig. \ref{fig:Results_Denoising_Art_Scene} shows better performance of the proposed BU3D algorithm when compared to classical maximum likelihood, and  Lindell's  \cite{LLindell_SigGraph2018} algorithms. Initial results on robust super-resolution by BU3D are also shown in Fig. \ref{fig:Results_SR} when considering an upsampling factor  $r=4$. Quantitative results and performance on real data are detailed in \cite{koo2022bayesian}.

\section{Conclusions}
This paper has presented an unrolling method  for joint depth reconstruction and super-resolution.  
We design our neural network by unrolling a
previous iterative Bayesian method \cite{halimi2021robust}, exploiting the domain
knowledge on a single-photon Lidar system. This unrolling
strategy makes the proposed network interpretable by the
connection to the Bayesian method and efficient in terms of the
network size, and the training and testing times. The resulting
network is robust to mismodeling effects due to 
differences between training and testing data as shown in \cite{koo2022bayesian}, and can be easily extended to perform super-resolution as indicated in this short paper.  
The numerical experiments show that the proposed
model can reconstruct high quality depth maps in challenging
scenarios with less artifacts around the surface boundaries.
Extending the model by accounting for the reflectivity maps
as input is interesting and will be studied in the future.

\textit{\footnotesize{This work was supported by the UK RAEng  Research Fellowship Scheme (RF/201718/17128) and EPSRC Grants EP/T00097X/1,EP/S026428/1.}}
 
 \newpage

\bibliographystyle{IEEEtran}
\bibliography{references.bib}
\end{document}